\documentclass[twocolumn,showpacs,nonbibnotes,nofootinbib]{revtex4-1}
\usepackage[T2A]{fontenc}
\usepackage[cp1251]{inputenc}
\usepackage[english]{babel}
\usepackage{amsmath}
\usepackage{amsfonts}
\usepackage{amssymb}
\usepackage{color}
\usepackage{makeidx}
\usepackage{hyperref}
\usepackage[pdftex]{graphicx}

\usepackage{bm}

\newcommand{\e}{\mathrm{e}}

\renewcommand{\d}{\mathrm d}

\newcount\timehh  \newcount\timemm
\timehh=\time \divide\timehh by 60
\timemm=\time
\begin{document}

\title{Effect of exchange interaction on the spin fluctuations of localized electrons}

\author{D.S. Smirnov}\email{Corresponding author: smirnov@mail.ioffe.ru}
\author{M.M. Glazov}\thanks{Also at: Spin and Optics Laboratory, St.-Petersburg State University, 198504
St.-Petersburg, Russia}
\author{E.L. Ivchenko}
\affiliation{Ioffe Physical-Technical Institute of the RAS, 194021 St.-Petersburg, Russia}

\begin{abstract}
In this paper a microscopic theory of spin fluctuations in an ensemble of electrons localized on donors in a bulk semiconductor has been developed. Both the hyperfine interaction of the electron spin with spins of lattice nuclei and the exchange interaction between the electrons have been taken into account. We propose a model of clusters to calculate spin noise spectra of the ensemble of localized charge carriers. It has been shown that the electron-electron exchange interaction leads to an effective averaging of random nuclear fields and a shift of the peak in the spin-fluctuation spectrum towards lower frequencies.
\end{abstract}

\maketitle

\section{Introduction}\label{sec:intro}
The rapid development of semiconductor spintronics and search for systems with very long spin relaxation times has led to the development of experimental techniques for spin dynamics studies. One of the most promising methods of this type is the spin noise spectroscopy, proposed by Aleksandrov and Zapasskii more than thirty years ago for investigation of magnetic resonances in atomic gases~\cite{aleksandrov_zapasskiy_noise:rus} (see also \cite{Crooker_Noise}), and developed in the last years to study dynamics of spins in bulk semiconductors, quantum wells and quantum dots~\cite{Oestreich_noise,muller:206601,PhysRevB.79.035208,crooker2010,crooker2012,dahbashi:031906,Zapasskii-condMat}, see also reviews~\cite{Mueller2010,1742-6596-324-1-012002}. Along with this, the spin noise of the nuclei~\cite{Reilly:2008vn} and fluctuations of spin current~\cite{Moca:2010uq, Moca:2011kx} are investigated experimentally and theoretically.

The spin noise spectroscopy is based on the transmission of a linearly-polarized probe beam through the sample and measurement of the fluctuations of the Faraday or Kerr rotation angle as well as of the ellipticity. These fluctuations are characterized by correlation functions which are related to the autocorrelation function of the total-spin components $\langle S_z(t)S_z(t') \rangle $, where $z$ is the  propagation direction of the probe beam. The Fourier transform of spin correlator characterizes the intensity of spin fluctuations, it contains information about the distribution of spin precession frequencies and spin relaxation and/or dephasing times.

The theory of spin fluctuations in ensembles of electrons and holes localized in quantum dots has been developed in Ref.~\cite{gi2012noise}. In particular, it is shown that the spin noise spectrum reveals peculiar properties of the distribution of random nuclear fields which play a major role in the dephasing of the localized charge carrier spins. Another promising system for the spin noise investigation is an ensemble of donors or acceptors in bulk semiconductors~\cite{Romer2010}. At low temperatures and moderate impurity densities, electrons and holes are localized and their spins effectively interact with spins of the lattice nuclei. These are those systems where extremely long electron spin relaxation times can be achieved~\cite{dzhioev97, Kikkawa98, Dzhioev02}. However, in bulk doped semiconductors the exchange interaction between localized electrons can play an important role~\cite{Dzhioev02, KKavokin-review}. The goal of this work is to investigate theoretically the manifestations of the exchange interaction between electrons in the spin 
fluctuations spectra.

\section{Model}\label{sec:model}

We consider an ensemble of electrons localized on donors in a GaAs-type bulk semiconductor and take into account the hyperfine interaction of electron spin with surrounding nuclear spins as well as the exchange interaction between localized electrons. The Hamiltonian of the system can be presented as follows
\begin{equation}
\label{Ham}
 {\hat{\mathcal{H}}}=\hbar\sum_{i}\mathbf{\Omega}_i\hat{\mathbf{s}}_i +\sum_{i\ne k}J_{ik}\hat{\mathbf{s}}_i\hat{\mathbf{s}}_k.
\end{equation}
Here the subscripts $i,k $ enumerate donors, $\hat{\mathbf{s}}_i$ is the electron spin operator with the components $\sigma_x^{(i)}/2,\sigma_y^{(i)}/2,\sigma_z^{(i)}/2 $, where $\sigma^{(i)}_\alpha$ ($\alpha = x, y, z $) are the Pauli matrices acting on the spin variables of the $i$-th electron (more precisely, of the electron localized on the $i$-th donor), $\mathbf{\Omega}_i $ is the electron spin precession frequency in the field of nuclear fluctuation, $J_{ik}=J_{ki}$ is the constant of exchange interaction between $i$-th and $k$-th electrons. The latter exponentially depends on the distance between donors $R_{ik}$~\cite{GP63, Herring:1964ys}:
\begin{equation}
{J_{ik}}=0.82 \mathcal R \left(\frac{R_{ik}}{a_B}\right)^{5/2}\exp\left(-2\frac{R_{ik}}{a_B}\right).\label{JR}
\end{equation}
Here $\mathcal R$ and $a_B$ are the electron Rydberg and the Bohr radius, respectively (in GaAs $\mathcal R \approx 4$~meV, $a_B\approx 120$~\AA), and it is assumed that $R_{ik} \gg a_B$. In equilibrium, nuclear spin fluctuations are described by a Gaussian distribution~\cite{Dzhioev:2002kx, gi2012noise}
\begin{equation}
\label{Gauss}
\mathcal F_{\delta_e}(\mathbf \Omega) = \frac{1}{\pi^{3/2} \delta_e^3} \exp{\left(-\frac{\Omega^2}{\delta_e^2}\right)},
\end{equation} 
where the parameter $\delta_e$ characterizes the fluctuation dispersion: $\langle\Omega^2\rangle = 3\delta_e^2/2$.
A typical value of the spin precession frequency in the field of nuclear fluctuation in Gallium Arsenide amounts to $\sqrt{\langle\Omega^2\rangle}\sim 2\times 10^8$~s$^{-1}$~\cite{Dzhioev:2002kx, Romer2010} so that, at the interdonor distance $\sim0.1~\mu$m, the electron-electron exchange interaction is comparable with the hyperfine interaction. This corresponds to the donor concentration $n_d=10^{14}\div10^{15}$~cm$^{-3}$. Note that in GaAs the ``metal-insulator'' transition occurs at much higher concentrations of donors $\sim2\times10^{16}$~cm$^{-3}$~\cite{Dzhioev02}. Thus, in order to describe the spin noise of electrons localized on donors in bulk semiconductors even at rather low level of doping, an interplay of the exchange interaction between electrons and the hyperfine interaction of electron and nuclear spins should be taken into account. The main aim of this work is to analyze this interplay.

Detection of spin fluctuations is usually performed by measuring the fluctuations of spin Faraday, Kerr and ellipticity signals~\cite{Crooker_Noise, crooker2012, muller:206601}.
Let the probe beam be focused into a spot of the area $S$ on the sample surface, the sample thickness be $L$ and the carrier frequency of the probe beam, $\omega_0$, lie slightly below the fundamental absorption edge frequency $E_g/\hbar$ ($E_g$ is the band gap) so that the beam propagates in the region of weak absorption. Instantaneous values of the Faraday or Kerr rotation as well as the ellipticity are determined by fluctuations of the total spin $\hat{\mathbf S}=\sum_{i=1}^N\hat{\mathbf s}_i$ of the subsystem of $N=n_d L S$ electrons being present in the volume illuminated by the probe beam. In accordance with this we introduce the correlation function $\langle\hat{S}_\alpha(t+\tau)\hat{S}_\beta(t)\rangle$ of the total spin components of this subsystem, where the angular brackets denote averaging over time $t$ with the fixed time difference $\tau$~\cite{ivchenko73fluct, springerlink:10.1007/BF02724353, ll5, gi2012noise}.
It is also convenient to introduce the correlation function $\langle{s}_\alpha(t+\tau){s}_\beta(t)\rangle$ and the spectral intensity $(s_\alpha s_\beta)_\omega$ of spin fluctuations normalized per particle, as follows
\begin{equation}
\label{single:corr}
\langle {s}_\alpha(t+\tau) {s}_\beta(t)\rangle = \frac{1}{N} \langle \hat{S}_\alpha(t+\tau) \hat{S}_\beta(t)\rangle, 
\end{equation}
\[
(s_\alpha s_\beta)_\omega = \int_{-\infty}^{\infty}\left\langle s_\alpha(t+\tau) s_\beta(t)\right\rangle\e^{i\omega \tau}\d\tau.
\]
Bellow we present the calculation of the above correlation functions.
 
We introduce the basis states $|n\rangle$, $|m\rangle$, etc., corresponding to the eigenvalues $E_n$, $E_m$, \ldots of the Hamiltonian \eqref{Ham}. By using the Heisenberg representation of the total spin operator, we obtain~\cite{Correlators}
\begin{multline}
 (s_\alpha s_\beta)_\omega= \\
 \frac{2^{1-N}\pi}{N} \sum_{n,m}
\langle n |\hat S_{\alpha} |m\rangle\langle m |\hat S_{\beta}|n\rangle \Delta\left(\omega - \frac{E_n - E_m}{\hbar}\right).
\label{HugeFormula}
\end{multline}
Here the summation is carried out over all eigenstates of the system, the broadened $\delta$-function
\[
 \Delta(x)=\frac{1}{\pi}\frac{\tau_s}{1+(x\tau_s)^2},
 \] 
is introduced with $\tau_s$ being the spin relaxation time of a single electron unrelated to the hyperfine or exchange interaction, considered as a phenomenological parameter of the theory and assumed to be the same for all the electrons. While deriving Eq.~\eqref{HugeFormula} it is assumed that the temperature of the system $T$ expressed in the energy units exceeds by far the characteristic interlevel energy splittings $|E_n - E_m|$. Hence, the averaging in Eq.~\eqref{HugeFormula} is performed with the equilibrium density matrix representing the equally occupied eigenstates. This condition is well satisfied in experiments on spin noise spectroscopy carried out at temperatures down to that of liquid Helium because the characteristic splittings between the $n$ and $m$ levels, due to both exchange and hyperfine interactions, correspond to $T\sim 10^{-3}$~K.

 \section{Spin noise in the model of clusters}\label{sec:model_clust}

The straightforward calculation of the spin noise spectra according to the general formula \eqref{HugeFormula} seems impossible because, under typical experimental conditions with the donor concentration amounting $n_d=10^{14}$~cm$^{-3}$, the probing spot area $S>1000$~$\mu$m$^2$, and the thickness of the probed layer $L>10$~$\mu$m, the number of donors in the probe volume exceeds $10^6$. However, the exponential dependence of exchange interaction constant $J_{ik}$ on the distance between the donors makes negligible the exchange interaction between  sufficiently distant electrons. Therefore, the exchange interaction is important only for a group of donors located close enough to each other. This allows one, for the calculation of spin noise spectra, to develop an approach similar to the percolation theory used to analyze effects of electron transport in disordered systems~\cite{Efros89}.

The structure of spin states in the ensemble of electrons is determined by the interplay of the exchange interaction between charge carriers and their hyperfine interaction with the lattice nuclei. Clearly, if the absolute values of nuclear fields $|\bm \Omega_i|$ and $|\bm \Omega_k|$ acting on the electrons $i$ and $k$ exceed the exchange interaction constant $J_{ik}$, then the exchange interaction between these electrons is inessential. Otherwise, it is the electron-electron exchange interaction which dominates, and the spin states of electrons are quantized according to the total magnetic momentum. In the frame of this model it is natural to decouple the whole ensemble of localized electrons into groups or clusters of donors. Within each cluster the exchange interaction dominates over the hyperfine interaction:
\begin{equation}
\label{crit}
J_{ik} > A\hbar \delta_e,
\end{equation}
where $A$ is a dimensionless coefficient of the order of unity, its precise value weakly affects the final results. Let us introduce the characteristic distance between the donors $R_c$ defined by $J(R_c) = A \hbar\delta_e$. Then the problem of cluster formation in the ensemble of electrons localized on donors is quite similar to that of formation of clusters in the studies of high-frequency conductivity of a disordered system. In the latter case, the parameter $R_c$ depends on the frequency of electromagnetic field and describes the displacement of an electron during the half-period of the field oscillation~\cite{achopping}. Strictly speaking, a particular cluster contains a group of donors with the nearest neighbour distances smaller than or equal to $R_c$. Let us introduce the dimensionless parameter
\begin{equation}
  \eta=\frac{\pi}{6}  n_d R_c^3,
\label{eta}
\end{equation} 
which shows the volume fraction of spheres of radius $R_c/2$. The parameter $\eta$ determines the statistics of clusters in the system, i.e. the function $\mathcal P(N)$ describing the probability to find a cluster made of $N$ electrons. We present the first two values of $\mathcal P(N)$ for the random distribution of donors~\cite{wisc}
\begin{eqnarray}
\mathcal P(1) &=& \exp{\left( - \frac{4 \pi n_d R_c^3}{3} \right)} = {\rm e}^{- 8 \eta}\:,\\
\mathcal P(2) &=& {\rm e}^{-8 \eta} \int\limits_0^{R_c}
4 \pi n_d r^2 dr \exp{\left[- \pi n_d \left( R_c^2 r - \frac{r^3}{12}\right) \right]} \nonumber\\
&=&24 \eta {\rm e}^{-8 \eta} \int\limits_0^1 u^2 \exp{\left[ - 6 \eta \left( u - \frac{u^3}{12} \right)\right]} du\:. 
\end{eqnarray}
The last integral monotonically decreases from $1/3$ to $\approx 0.1$ as $\eta$ varies between $0$ and $0.2$.

Note that, for $\eta \approx 0.34$, an infinite cluster is formed in the system~\cite{Efros89, Threshold}. This corresponds to a critical concentration of donors $n_c\sim 10^{15}$~cm$^{-3}$. We focus on the case of $n_d\ll n_c$ where the vast majority of clusters are represented by single electrons and groups of a small number of donors, $N=2,3,4$, with strong exchange interaction between them. Even at $\eta=0.1$ the probability to find a cluster composed of four electrons is less than $8\%$, and the values of $\mathcal{P}(N)$ for $N > 4$ are negligibly small.

The level structure in the cluster consisting of $N$ electrons is determined primarily by the exchange interaction. Consequently, in the zeroth-order approximation in the hyperfine interaction, the available $2^N$ levels are combined into groups related to a fixed total spin $M$ of all electrons (for even $N$ the total spin of electrons can be $0,1,\ldots$ up to $N/2 $, and for odd $N$ it is $M=1/2,3/2,\ldots N/2$). The number of groups of states (or multiplets) with the spin $M$ is given by~\cite{ll3}
\begin{equation}
\label{NM}
\mathcal N(M) = \frac{(2M+1)N!}{(N/2 + M +1)!(N/2-M)!}.
\end{equation}
Let us introduce the electron eigenfunctions $\Psi_l(M, m)$ and eigenenergies $E_l(M)$ in the cluster, where $m$ is a projection of the total spin of $M$ on the $z$ axis running through $2M+1$ values from $-M$ to $M$, the index $l$ enumerates $\mathcal N(M)$ different energy levels with a given value of the total spin $M$. The energy gaps between these levels are determined by the exchange interaction constants. Phases of the eigenfunctions are chosen in such a way that the sets $\Psi_l(M,m)$ with fixed $l$ and $M$ transform under coordinate transformations in the spin space according to the representation $D_M$ as the spherical functions $Y_{Mm}(\theta,\varphi)$. In this case the
matrix elements of the total spin operator ${\bm S}_N=\sum_{i = 1}^N{\bm s}_i$ in the cluster are determined by
\[
\langle {\Psi_l(M,m')} | S_{\alpha} | {\Psi_l(M,m)} \rangle = J^M_{\alpha; m'm}\:,
\]
where $J^M_{\alpha; m'm}$ are the standard matrices of the operators representing components of the angular momentum $M$. Since the direct product
$$
D_{M} \times D_{M} = \sum\limits_{M' = 0}^{2M} D_{M'}
$$
contains only one representation $D_1$, the matrix elements of the $i$-th individual spin are proportional to the components of matrix $J^M_{\alpha;m'm}$, namely,
\begin{equation} \label{siJ}
\langle {\Psi_l(M,m')} | s_{i,\alpha} | {\Psi_l(M,m)} \rangle = c_i^{(M,l)} J^M_{\alpha; m'm}\:.
\end{equation}
Here $c_i^{(M,l)}$ are coefficients dependent solely on the energy level $l$ but not on $m$ and $m'$. We decompose the state $\Psi_l(M,m)$ as follows
\begin{equation} \label{siJ2}
\Psi_l(M,m) = \sum\limits_{m_1\dots m_i \dots m_N} C^{(M,m,l)}_{m_1 \dots m_i \dots m_N} {\chi}(m_1 \dots m_i  \dots m_N)\:,
\end{equation}
where $m_i=\pm 1/2$ is the spin projection of the $i$-th electron on the axis $z$, $i = 1,2\dots N$, $\chi(m_1 \dots m_i \dots m_N)$ is an $N$-particle basis spin function of the state with a given set of spin projections $m_i$. Apparently, the coefficients $C^{(M,m,l)}_{m_1\dots m_i\dots m_N}$ can differ from zero only if $\sum_i m_i = m$. The coefficient $c_i^{(M, l)}$ in Eq.~\eqref{siJ} is expressed via the coefficients of the expansion Eq.~\eqref{siJ2} as
\begin{multline}
c_i^{(M,l)} =
 \sum\limits_{m_1 \dots m_{i-1} m_{i+1} \dots m_N} \frac{1}{2M}  \\
 \times \left(  \vert C^{(M,m,l)}_{m_1 \dots m_i=1/2 \dots m_N} \vert^2 - \vert C^{(M,m,l)}_{m_1 \dots m_i=-1/2 \dots m_N} \vert^2 \right) \:.\nonumber
\end{multline}
The values of $C^{(M,m,l)}_{m_1 \dots m_i\dots m_N}$ and, hence, $c_i^{(M, l)}$ depend on the particular realization of exchange interaction constants in the cluster and, in the general case, can be found only numerically.

In the first approximation in the hyperfine interaction, different multiplets, i.e. states with different values of $M$ and $l$, are unmixed, but within each group with given $M$ and $l$ the states are split into $2M+1$ sublevels due to interaction of spins of electrons and nuclei. The effective Hamiltonian of hyperfine interaction can be presented as
\begin{equation}
\label{Hml}
\hat{\mathcal H}^{(M,l)} = \hbar \hat{\mathbf M}_l\bm \Omega_{\rm eff}^{(M,l)},
\end{equation}
where $\bm \Omega_{\rm eff}^{(M,l)}$ is an effective nuclear field acting on the total spin $\mathbf M_l$. According to Eqs.~\eqref{Ham} and~\eqref{siJ} it has the form
\begin{equation}
\label{Omega:eff}
\bm \Omega_{\rm eff}^{(M,l)} = \sum_i c_i^{(M,l)} \bm \Omega_i.
\end{equation}
The hyperfine interaction completely lifts the degeneracy in the projection of total spin $M$ and splits the degenerate level $E_l(M)$ into equidistant sublevels $E_l(M,M_{z '})$ with the given projections $M_{z'}$ of the total angular momentum on the axis of the effective nuclear field, and the splitting $\hbar\Omega_{\rm eff}^{(M,l)}$ between the neighboring sublevels. We denote the corresponding eigenfunctions of the system by $\Psi'_l(M,M_{z'})$. It should be emphasized that the direction of the $z'$ axis depends both on particular realization of nuclear fields and the parameters $M$, $l$ characterizing realization of the state with a given total spin of electrons in the cluster.

The spin noise spectrum of an ensemble of electrons in the cluster model is a sum of contributions from different clusters and can be written as
\begin{equation}
 (s_\alpha s_\beta)_\omega= \sum_{N=1}^\infty \mathcal P(N)  \sum_{M} \sum_{l=1}^{\mathcal N(M)} \mathcal S_{\alpha\beta}(N,M,l;\omega),
\label{HugeFormula1}
\end{equation}
where $\mathcal S_{\alpha\beta}(N,M,l;\omega)$ is a partial contribution to the spin noise of the $l$-th realization of total spin $M$ in the cluster of $N$ electrons [cf. with Eq.~\eqref{HugeFormula}]:
\begin{multline}
\label{SNMl}
\mathcal S_{\alpha\beta}(N,M,l;\omega) = \frac{2^{1-N}\pi }{N} \\
\sum_{M_{z'}, M_{z'}'} 
\langle \Psi'_l(M,M_{z'}) |\hat M_{l,\alpha} |\Psi'_l(M,M'_{z'})\rangle \\ 
\times \langle \Psi'_l(M,M'_{z'}) |\hat M_{l,\beta}|\Psi'_l(M,M_{z'})\rangle \\ \times \Delta\left(\omega - \frac{E_l(M,M_{z'}) - E_l(M,M_{z'}')}{\hbar}\right).
\end{multline}
In Eq.~\eqref{SNMl} the operators $\hat{M}_{l,\alpha}$, $\hat{M}_{l,\beta}$ change the spin projection by no more than~$1$, therefore $M_{z'}' = M_{z'}, M_{z'} \pm 1$ so that the argument of broadened $\delta$-function can be $\omega$ or $\omega \pm \Omega_{\rm eff}^{(M,l)}$. Here the averaging over all possible realizations of the nuclear fields acting on the localized electrons is assumed, whereupon the off-diagonal components of $\mathcal S_{\alpha\beta}$ vanish and the diagonal ones become equal, $\mathcal S_{xx}=\mathcal S_{yy}=\mathcal S_{zz} \equiv \mathcal S$~\cite{gi2012noise}. Finally, we obtain
\begin{multline}
\label{SNMl:fin}
\mathcal S(N,M,l;\omega) = \frac{2^{1-N}\pi}{9N} M(M+1)(2M+1)  \biggl\{ \Delta(\omega)   \\
\left. + \int \mathrm d {\bm{\Omega}}_{\rm eff} \mathcal F_{\delta_{M,l}}(\mathbf{\Omega}_{\rm eff}) \left[\Delta(\omega - \Omega_{\rm eff}) + \Delta(\omega + \Omega_{\rm eff})\right] \right\}.
\end{multline}
While deriving Eq.~\eqref{SNMl:fin}, it is taken into account that, for independent Gaussian distributions of spin precession frequencies $\mathbf \Omega_i$, their weighted sum \eqref{Omega:eff} is also distributed according to the Gaussian Eq.~\eqref{Gauss}, with the replacement $\delta_e \to \delta_{M,l}$, where the parameter $\delta_{M,l}$ is determined by the specific realization of the coefficients $c_i^{(M,l)}$~\cite{feller}:
\begin{equation} \label{dMl}
 \delta_{M,l} = \delta_e\sqrt{\sum_i \left(c_i^{(M,l)}\right)^2}\:.
\end{equation}
For the total spin $M = 0$, the spectral function \eqref{SNMl:fin} is identically zero, because the nuclear field does not act on the singlet state and the mean square of spin fluctuations in the singlet state is zero.

Equation~\eqref{SNMl:fin} for the contribution of the $l$-th realization of the total spin $M$ to the spin fluctuation spectrum of the cluster can be obtained without direct computation of matrix elements of operators of angular momentum $\hat{M}_{l,\alpha}$, $\hat{M}_{l,\beta}$. It can be done instead by using the method of Langevin random forces developed in \cite{gi2012noise} for fluctuation $\delta\mathbf M$ of the total spin of the ensemble. According to Eq.~\eqref{Hml} the fluctuation $\delta\mathbf M$ satisfies the Bloch equation
\begin{equation}
\label{fluct1:eq}
 \frac{\d \delta \mathbf{M}}{\d t}+\delta \mathbf{M}\times\bm \Omega_{\rm eff}^{(M,l)} + \frac{\delta \mathbf M}{\tau_s}=\bm \Xi(t),
\end{equation} 
whose right-hand side contains a fictitious random force $\bm\Xi(t)$ with the correlator
\begin{equation}
\label{langevin1}
\langle \Xi_\alpha (t) \Xi_\beta(t')\rangle = \frac{2}{3\tau_s}M(M+1) \delta_{\alpha\beta} \delta(t-t').
\end{equation}
Solving Eq.~\eqref{fluct1:eq} and calculating the correlation functions $\langle\delta M_\alpha(t)\delta M_\beta(t ')\rangle$ averaged over the distribution of fields $\bm\Omega_{\rm eff}^{(M, l)}$, we arrive at Eq.~\eqref{SNMl:fin}.

\section{Approximation of an ensemble of pairs and singles}\label{sec:2}

The influence of hyperfine and exchange interactions on spin fluctuation spectrum of localized electrons is most clearly traced for a pair of closely spaced donors $1$ and $2$ so that the exchange interaction with other donors can be neglected: ${J\equiv}J_{12}\gg J_{1i}, J_{2{i}}$ (${i}\ne 1,2$). In this section, we analyze this particular case.

\begin{figure}[t]
\includegraphics[width=\linewidth]{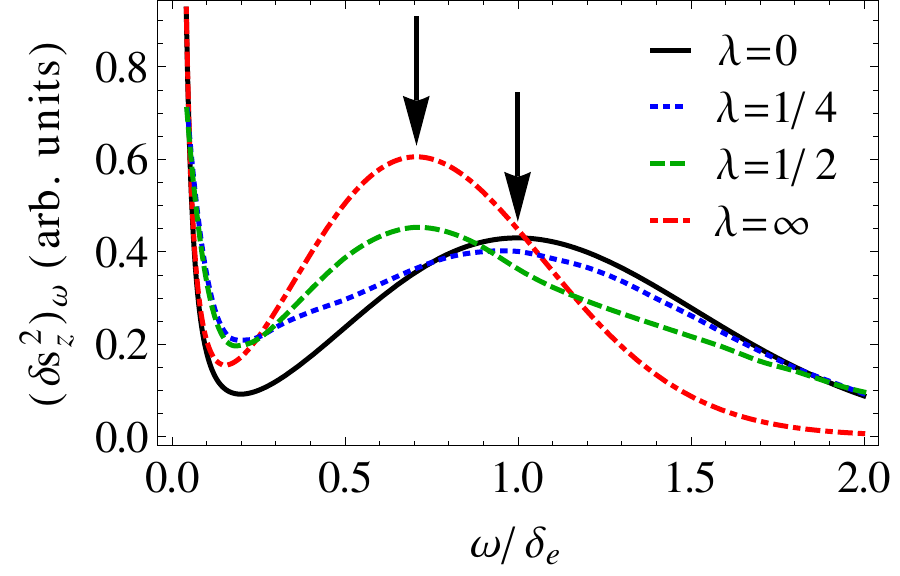}
\caption{ Spin fluctuation spectrum of a pair of electrons calculated for different values of the parameter ${\lambda\equiv}J/(\hbar\delta_e)=0$ (solid line), $1/4$ (dots), $1/2$ (dashed) and in the limit $\lambda\to\infty$ (dot-dashed). Arrows indicate the positions of the peak of spin noise in the limiting cases of weak ($\omega=\delta_e$) and strong ($\omega=\delta_e/\sqrt{2}$) exchange interaction.}
\label{numeric}
\end{figure}

In the limit of weak exchange interaction (${J}\ll\hbar\delta_e$) electrons can be considered as non-interacting, and the spin fluctuation spectrum has the form~\cite{gi2012noise} [cf. with Eq.~\eqref{SNMl:fin}]:
\begin{multline}
  (\delta s_z^2)_\omega=\mathcal S(1,1/2,1;\omega) = \\
  \frac{\pi}{6}\left\lbrace\Delta(\omega)+ \int \d\mathbf{\Omega}\mathcal{F}_{\delta_e}(\mathbf{\Omega})\left[\Delta(\omega-\Omega)+\Delta(\omega+\Omega)\right]\right\rbrace.
  \label{scl}
\end{multline} 
The spectrum calculated after this equation is shown by the dark solid curve in Fig.~\ref{numeric}. At $\omega\geqslant 0$ it consists of two peaks, the first of them is centered at $\omega = 0$ and associated with the fluctuations of spin component parallel to the nuclear field, while the second peak, centered at $\omega = \delta_e$, is related to the spin precession in the nuclear field, its shape is described by the distribution function of the absolute values of nuclear fields, $F_{\delta_e}(\Omega) = 4\pi\Omega^{2}\mathcal{F}_{\delta_e}(\Omega)$. Hereafter we consider a realistic case $\delta_e\tau_s\gg 1$. Note that the spectrum \eqref{scl} coincides with the first spectral function $\mathcal S_{\alpha\beta}(1,1/2,1;\omega)$ in Eq.~\eqref{HugeFormula1}, since a single (isolated) donor corresponds to $N=1$, the electron spin $1/2$ and the single realization $l=1$.

Now we turn to the consideration of the limiting case of strong exchange interaction in the pair of localized electrons: $J\gg\hbar\delta_e$. Neglecting the hyperfine interaction, the states of a pair of electrons are characterized by the pair's total spin $S=0$ (singlet) or $S=1$ (triplet) and the spin projection $S_z$ on a given axis $z$. If the splitting between the singlet and triplet is large, one can neglect mixing of these states due to the hyperfine interaction and it suffices to consider the dynamics of the triplet state with total spin $S=1$. The fluctuations of the triplet spin $\delta\mathbf{S}$ are described by Eq.~\eqref{fluct1:eq} with the effective nuclear field
\begin{equation}
\label{omega:triplet}
{\mathbf \Omega}_{\rm eff}=\frac{\mathbf \Omega_1 + \mathbf \Omega_2}{2}.
\end{equation} 
In accordance with Eq.~\eqref{dMl} the dispersion of ${\mathbf\Omega}_{\rm eff}$ is by a factor of $\sqrt{2}$ smaller than the dispersion of fields $\bm\Omega_1$, $\bm\Omega_2$ acting on the individual electrons. At the same time the spin noise spectrum per electron is determined by Eq.~\eqref{scl} with the replacement of $\mathcal F_{\delta_e}(\mathbf\Omega)$ by $\mathcal F_{\delta_e/\sqrt{2}}(\mathbf\Omega)$. The spin noise spectrum in this limiting case is shown by the red dot-dashed line in Fig.~\ref{numeric}. One can see from the figure that the peak at $\omega=0$ is the same as that obtained neglecting the exchange interaction, while maximum of the peak related to the spin precession is shifted to the frequency $\delta_e/\sqrt{2}$, the peak height is $\sqrt{2}$ times larger, and its width is respectively $\sqrt{2}$ times smaller. This occurs due to the effective averaging of nuclear fields caused by the exchange interaction. The dot-dashed curve in Fig.~\ref{numeric} reveals the spectral function $\mathcal S_{\alpha\beta}(2,1,1;\omega)$ in the expansion Eq.~\eqref{HugeFormula1} of the cluster model.

In the framework of cluster model described above, in a cluster of two donors the exchange constant $J$ exceeds the hyperfine splitting $\hbar\delta_e$. For two donors one can apply the general formula~\eqref{HugeFormula} and trace the evolution of spin noise spectrum with the continuous change from zero to infinity of the parameter $\lambda=J/(\hbar\delta_e)$ characterizing the relative strength of the exchange and hyperfine interactions. In addition to the above-mentioned limiting curves calculated for $\lambda=0$ and $\lambda=\infty$, Fig.~\ref{numeric} shows the spectra corresponding to $\lambda = 1/4$ and $\lambda = 1/2$. Qualitatively the spectral shape is the same for various values of $\lambda$. However, as $\lambda$ increases the peak related to the spin precession shifts towards the lower frequencies.
For small values of $\lambda$ the spin precession peak is centered at $\omega=\delta_e$, and in the limiting case of strong exchange interaction the peak shifts to $\omega=\delta_e/\sqrt{2}$. The switching between the two regimes under changing $\lambda$ occurs at $\lambda\approx 0.33$. The transition is quite abrupt so that, for $\lambda=0.25$ and $0.5$, the peak positions take the limiting values. The abrupt character of the transition confirms the applicability of cluster model, and the transitional value of $\lambda$ shows that, in the inequality \eqref{crit}, the value of $A=1/3$ can be considered as optimal.

\begin{figure}[t]
\includegraphics[width=\linewidth]{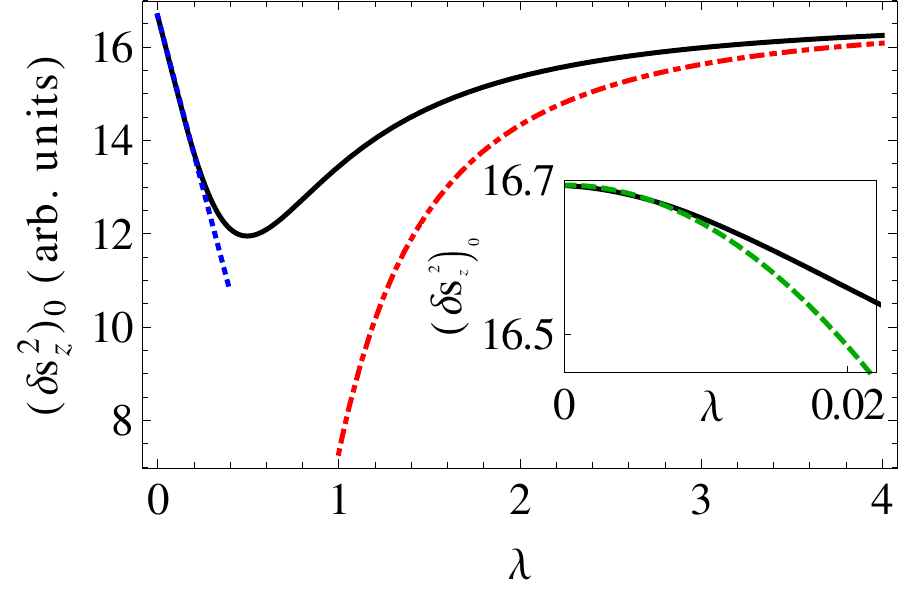}\\
\caption{Amplitude of the spin fluctuations at zero frequency. The solid curve is a result of numerical calculation after Eq.~\eqref{HugeFormula}; dotted, dot-dashed and dashed (in inset) curves show the asymptotics found using Eqs.~\eqref{app2}, \eqref{app3} and \eqref{app1}, respectively. The calculations are performed for $\delta_e\tau_s = 50$.}
\label{fig:noise0}
\end{figure}

Now we will analyze the spin fluctuations at low frequencies, $\omega\lesssim\tau_s^{-1}$, where the spectrum is determined by spin relaxation processes. One can see from the calculations presented in Fig.~\ref{numeric} that the exchange interaction does not lead to disappearance of the peak at zero frequency. This is due to the fact that the projection of the total spin on the direction of the average nuclear field $\mathbf{\Omega}_{\rm eff}$ exhibits no changes with time. Figure~\ref{fig:noise0} shows the amplitude of the peak at $\omega=0$ as a function of the ratio $\lambda=J/\hbar\delta_e$. The amplitudes of peak at zero frequency in the limiting cases $J\ll\hbar\delta_e$ and $J\gg\hbar\delta_e$ coincide, see Eq.~\eqref{scl}. For $\lambda\approx 0.5$ the spin noise amplitude at zero frequency reaches a minimum amounting to about $3/4$ of its limiting values.

Analytical expressions for the dependence of $(\delta s_z^2)_{\omega=0}$ on the parameter $\lambda$ in the limiting cases of weak and strong exchange interaction can be obtained from the analysis of temporal dynamics of the two-particle density matrix $\rho(t)$ of the system at $t\gg 1/\delta_e, \hbar/J$ satisfying the equation
 \begin{equation}
 \label{dens}
  \frac{\d \rho(t)}{\d t}=\mathrm i[\rho(t),\mathcal{H}] -\mathcal L\{\rho(t)\},
 \end{equation} 
where $\mathcal{L}$ is a linear operator describing the spin relaxation with the characteristic time $\tau_s$. In the case of extremely weak exchange interaction $J\ll\hbar/\tau_s{\ll  \hbar \delta_e}$ we obtain
\begin{equation}
 (s_z^2)_0 \approx \frac{\tau_s}{6} \left( 1 + \frac{4}{(\delta_e\tau_s)^2} - \sqrt{\frac{\pi}{8}}{\tau_s\delta_e\lambda^2}
\right),
 \label{app1}
\end{equation} 
where, along with the correction related to the small parameters ${\lambda}$ and $J\tau_s/\hbar$, we take into account a small term proportional to $(\delta_e\tau_s)^{-2}$ which is independent of the exchange interaction and originating from nuclear fields (in the following formulas this term is omitted). A quadratic reduction of the amplitude of fluctuations with the increasing constant of exchange interaction is shown in the inset to Fig.~\ref{fig:noise0}. In the intermediate case when $\hbar/\tau_s\ll J\ll\hbar\delta_e$, one may restrict oneself only by terms of the first order in the exchange interaction constant,
\begin{equation}
 (s_z^2)_0\approx\frac{\tau_s}{6}\left(1 {- \frac{\sqrt{\pi}}{2}\lambda}\right),\label{app2}
\end{equation} 
and the amplitude of the peak at zero frequency decreases linearly with the increasing exchange interaction between localized electrons, see the dashed line in Fig.~\ref{fig:noise0}. In the opposite limit of $J\gg\hbar\delta_e$ we have the inverse square dependence of spin fluctuations amplitude on $\lambda$:
\begin{equation}
(s_z^2)_0\approx\frac{\tau_s}{6}\left[1 -\left({\frac{3}{4\lambda}}\right)^2\right]\:,\label{app3}
\end{equation} 
in agreement with numerical calculation (dot-dashed line in Fig.~\ref{fig:noise0}).

We conclude this section by a brief analysis of the role of spin-orbit interaction. Allowance for this interaction leads to an anisotropy of the exchange interaction in a pair of localized electrons which is now described by a tensor of exchange constants as follows
\begin{equation}
\label{exch:so}
J_{ij}^{\alpha\beta} \hat s_{i,\alpha} \hat s_{j,\beta}.
\end{equation}
The precise form of tensor $J_{ij}^{\alpha\beta}$ is determined by the symmetry of the system and the mechanism of spin-orbit interaction. Allowance for cubic in the wave vector terms in the effective Hamiltonian of a free electron (Dresselhaus effect)~\cite{KKavokin-review} gives rise to the relation between the components of $J_{ij}^{\alpha\beta}$ which allows one to rewrite Eq.~\eqref{exch:so} as $J_{ij}\hat s_i^{\prime}\hat s_j'$, where the operators $\hat s_i'$ and $\hat s_j'$ are connected with $\hat s_i$ and $\hat s_j $ by a unitary transformation, and the constants $J_ {ij}$ are again described by Eq.~\eqref{JR}. The singlet-triplet structure of states of the pair of electrons remains the same as in the absence of spin-orbit interaction. Neglecting the mixing of conduction and valence bands states, one can describe the spin noise spectrum of a pair of electrons averaged over the nuclear fields disregarding the spin-orbit coupling because the unitary transformation of the terms $\mathbf\Omega_i\hat{\mathbf s}_i $ leads only to a variation of the direction of the pseudovector $\mathbf\Omega_i$. However, allowance for the spin-dependent mixing of valence band states in the Kane model completely lifts the degeneracy of the spin states of the pair of electrons~\cite{glazov2009}.

\section{Beyond the approximation of pairs}\label{sec:clust}

\begin{figure}[hptb]
\includegraphics[width=\linewidth]{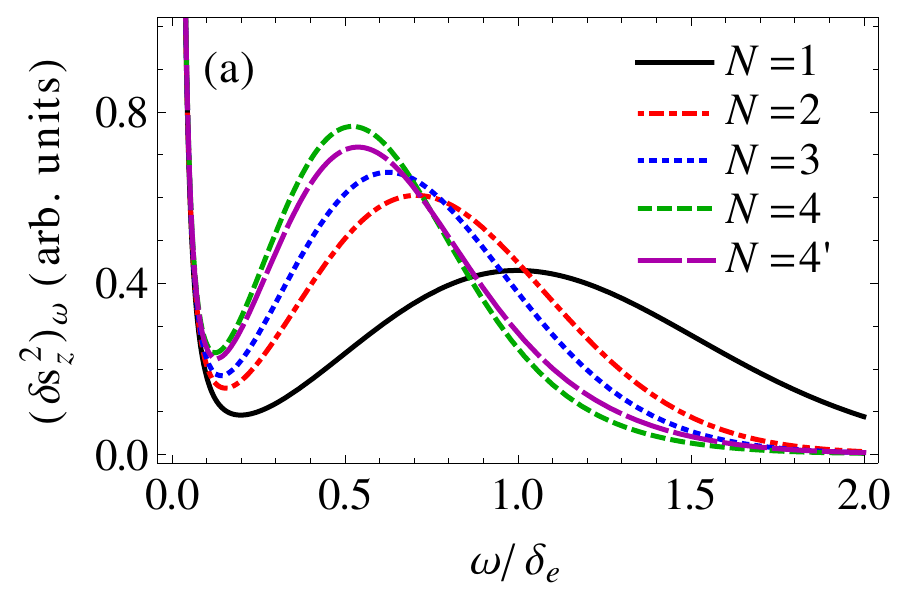} \\
\includegraphics[width=\linewidth]{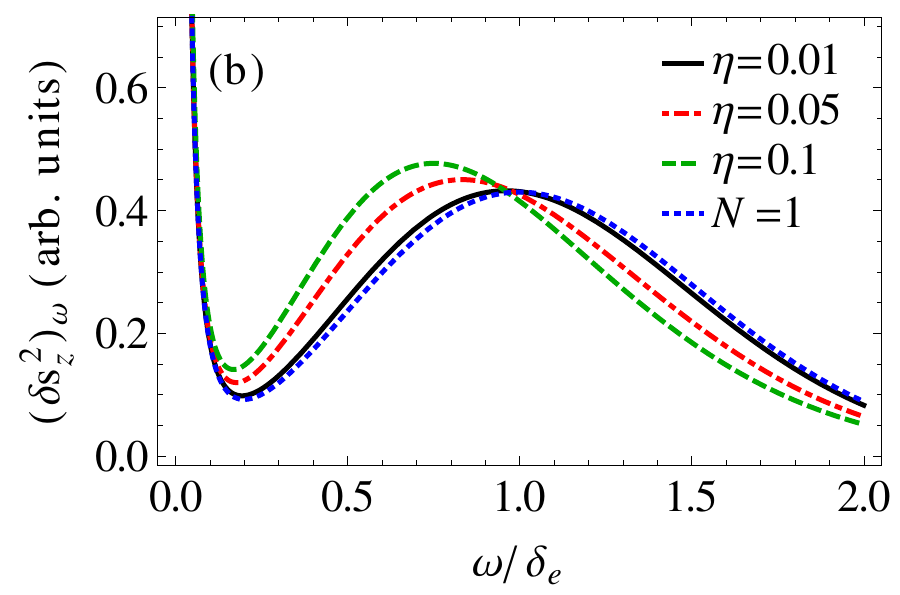}
\caption{
(a) The spin noise spectrum, normalized per electron, in a cluster of $ N $ donors. The curves marked with $ N = 4 $ and $ N = 4' $ (short and long dashes) differ by the relation between the exchange interaction constants, see text for details.
(b) Spin noise spectrum, normalized per electron, calculated in the cluster model for an ensemble of randomly distributed donors. Blue/dashed curve is obtained neglecting the exchange interaction and corresponds to a system of isolated donors. The calculations are performed for $\tau_s\delta_e = 100$.}\label{szN}
\end{figure}

Next we consider clusters with the number of donors $N>2$. An abrupt transition at ${\lambda}\approx 0.33$ between the limiting cases of weak and strong exchange interaction shown in Fig.~\ref{numeric} permits one to set $A = 1/3$ in Eq.~\eqref{crit}. For the calculation of the spin noise spectrum it is necessary to determine the half-widths $\delta_{M,l}$ introduced in Eq.~\eqref{dMl} which characterize the dispersion of nuclear fields acting on the $l$-th realization of the total spin $M$ in the cluster of $N$ donors. For three donors, the exchange interaction between electrons splits the eight-fold degenerate state into two sublevels $l = 1,2$ with $ M = 1/2 $ and one sublevel with $ M = 3/2 $. The corresponding parameters $\delta_{1/2,l}$ for $l = 1$ and $2$ are equal and the three half-widths $\delta_{M, l}$ are determined by the total electron spin: $\delta_{M, l} = \delta_e / \sqrt{2M} $, independent of the relation between exchange interaction constants $J_{12}, J_{23}$ and $J_{31}$. Moreover, the model calculation based on the general expression~\eqref{HugeFormula} in the case of three electrons shows that even for $J_{12},J_{23},J_{31}\approx\hbar\delta_e$ the fluctuation spectrum is close to that obtained from  Eqs.~\eqref{HugeFormula1}, \eqref{SNMl:fin} in the cluster model: the difference in vicinity of the peak related to the spin precession reaches no more than 30\%. Partial contributions to the spin noise spectrum from donor singles, pairs and triples calculated after Eq.~\eqref{SNMl:fin} are shown in Fig.~\ref{szN}(a) by solid, dot-dashed and dotted curves, respectively.

The situation for clusters with $N\geqslant4$ is more complicated, and the parameters $\delta_{M, l}$ depend on the relation between the exchange interaction constants. For example, for four donors in the cluster, there are three different realization $l = 1,2,3$ of the total spin $ M = 1 $, a single realization with $ M = 2 $, as well as two realizations with $ M = 0 $ (not contributing to the spin noise). If the exchange interaction in two pairs of the four dominates over interaction between the pairs, so that $ J_{12}, J_{34} \gg J_{13}, J_{14}, J_{23}, J_{24} $, then the relations $ \delta_{1,1} = \delta_{1,2} = \delta_e/\sqrt{2} $ and $\delta_{1,3} = \delta_e/2$ are valid. On the other hand, if $J_{12}\gg J_{13}, J_{23} \gg J_{14}, J_{24}, J_{34}$, then $\delta_{1,1}$ and $\delta_{1,2}$ remain unchanged whereas $ \delta_{1,3} = \delta_e\sqrt{7/12}$. For these two particular cases, the corresponding spin noise spectra are shown with short- and long-dashed lines in Fig.~\ref{szN}(a). The difference between them does not exceed 10\%.

Figure~\ref{szN}(b) presents the spin noise spectra of an ensemble of randomly distributed donors calculated in the cluster model. The values of $\mathcal P(N)$ for $N>2$ are calculated by means of the Monte Carlo method. To simplify the calculation, we assumed that all realizations of the same total spin $M$ experience the characteristic nuclear field $\delta_{M,l} = \delta_e/\sqrt{2M}$. At low donor densities, where $\eta = 0.01$ (solid curve), most clusters consist of single donors and the spin noise spectrum in fact coincides with that found from Eq.~\eqref{scl} for isolated centers (curve shown by blue dots). With the increase in donor concentration the spin fluctuation noise is determined by clusters of increasing size. Therefore, the peak related to the spin precession in the nuclear field shifts towards the lower frequencies and somewhat narrows down.

As noted above, at a sufficiently high concentration of donors where the dimensionless parameter $\eta$ reaches the value of 0.34, an infinite cluster of electrons interconnected by the strong exchange interaction appears in the system. In this case the model presented above is inapplicable, and the spin dynamics and spin fluctuations can be determined by an interference of spin-orbit and exchange interactions~\cite{KKavokin-review}. Moreover, if the probe beam spot area $S$ on the sample surface is small enough, $S\lesssim 10$~$\mu$m, then the shape of the spin noise spectrum can vary with the changing position of the illumination spot on the sample depending on whether the spot captures the clusters with large number of donors or not. Besides, the exchange interaction between electrons in the cluster can give rise to rather efficient spin diffusion from the illumination spot and the restriction imposed on the time $\tau_s\lesssim S/D$. Here, $D\sim n_d^{-2 /3}\langle J\rangle /\hbar$ is the spin diffusion coefficient, $ \langle J \rangle $ is the exchange constant ``averaged'' over the cluster. For example, for $n_d = 10^{16}$~cm$^{-3}$ and $ S = 10$~$\mu$m$^2$ the spin diffusion coefficient is $D\approx1.5$~cm$^{2}/$s and $S/D\sim7\times10^{-8}$~s which is comparable to the spin relaxation time caused by spin-orbit interaction~\cite{KKavokin-review}.

In the above we have discussed the fluctuations of total spin $\mathbf S$ of an ensemble of localized electrons. The exchange interaction between electrons in the absence of nuclear fields ($\mathbf\Omega_i = 0$) does not lead to a loss of the total spin of the ensemble but results in the peculiarities of spin dynamics of a single electron, which determines, e.g., a slow dynamics of nuclear spins~\cite{PhysRevB.81.115107}. In particular, for a pair of electrons with $\mathbf\Omega_1 = \mathbf \Omega_2 = 0 $ the asymptotic behavior of correlation functions $ \langle s_{i, \alpha} (t + \tau) s_{j, \beta} (t) \rangle $ in the limit $ \tau \to \infty $ turns out to be similar in the two limiting cases: in the absence of exchange interaction ($ J_{12} = 0 $):
\begin{subequations}
\begin{equation}
\label{J0}
\langle s_{i,\alpha}(t+\tau) s_{j,\beta}(t) \rangle =  \frac{1}{4} \delta_{ij} \delta_{\alpha\beta}  \mathrm e^{-\tau/\tau_s} \quad (\tau \gg \tau_s), 
\end{equation}
and in the presence of exchange interaction $J_{12} \gg \hbar/\tau_s$:
\begin{equation}
\label{Jm}
\langle s_{i,\alpha}(t+\tau) s_{j,\beta}(t) \rangle = \frac{1}{8}  \delta_{\alpha\beta}\mathrm e^{-\tau/\tau_s}  \quad (\tau \gg \tau_s). 
\end{equation}
\end{subequations}
The difference in prefactors can be explained if we take into account that the exchange interaction suppresses spin correlations of an individual electron and leads to an appearance of correlation between spins of different carriers. A similar behavior also takes place for clusters with larger numbers of electrons $ N $: due to the exchange interaction the spin is redistributed among all the electrons and, in the absence of nuclear fields, $ \langle s_{i, \alpha} (t + \tau) s_{i, \beta} (t) \rangle {\approx \delta_{\alpha \beta}} \exp {(- \tau / \tau_s)} / (4N) $. This effect is analogous to the influence of inter-electron interaction on the distribution function of a gas of free carriers: the electron-electron collisions conserve the total momentum of the gas but (i) result in the loss of momentum of an individual electron and (ii) lead to correlations of electron momenta after collision~\cite{springerlink:10.1007/BF02724353}. Features of temporal dynamics of the spin correlation function of an electron localized on a given center, with allowance for both the hyperfine interaction and exchange interaction with other electrons, can be analyzed using methods developed in~\cite{Feigelman:2010bh,Cuevas:2012cr} and require a separate study. Nevertheless, one can argue that, in the clusters with $J_{ij} \gg \hbar \delta_e $, the characteristic correlation time of the spin of a single electron has the order of $ \hbar / \langle J \rangle $.

\section{Conclusion}\label{sec:concl}

In the present work we have developed a theory of the spin fluctuations of localized electrons taking into account both the hyperfine interaction with lattice nuclei and the exchange interaction between localized charge carriers. The specific case of electrons localized on donors in an $ n $-type bulk semiconductor at low temperatures has been considered. The system has got three independent sources of randomness: (i) the random character of spin-flips during the spin relaxation described by the time $ \tau_s $, (ii) the random scatter of the three-dimensional vector of nuclear field acting on an electron localized on the $i$-th donor and, finally, (iii) the random distribution of donor centers in the sample which leads to a spread of exchange interaction constants. It has been shown that the spin noise spectrum of an ensemble of localized electrons has two peaks: the peak at $ \omega = 0 $ originates from fluctuations of the component of the total spin in the cluster of donors directed parallel to the total nuclear field, and the peak at $ \omega> 0 $ 
arises due to the precession of electron spins in the nuclear fields. The exact shape and position of this peak depend on the magnitude of nuclear fluctuations and strength of exchange interaction. With the increasing density of localization centers, the peak related to the spin precession narrows down and shifts toward the lower frequencies.

Modification of the spin noise spectrum with the application of an external magnetic field $ {\bm B} $ can be considered similarly to the approach of Ref.~\cite{gi2012noise} developed for isolated spins 1/2. The external field is added to the nuclear field and changes statistics of fluctuations of the total field acting on the spin of localized electron. If these fields have the comparable order of magnitude and, by definition of a cluster, are small compared to the exchange interaction in the cluster, one can neglect their influence on the splitting between the states with different values of $M$ and $l$ and, as before, consider the contribution of each multiplet to spin noise independently taking into account the splitting of its components in the first order of perturbation theory.

The authors are grateful to M.V. Feigel'man for useful discussions.

This work is partially supported by RFBR, the grant of the President of the Russian Federation NS-5442.2012.2, the Ministry of Education and Science (contract 11.G34.31.0067 with St. Petersburg State University and the leading scientist A.V. Kavokin), the Dynasty Foundation-ICFPM, and the EU grant SPANGL4Q.

\end{document}